\documentclass[prb,amsmath,amssymb,superscriptaddress,twocolumn]{revtex4-1}
\usepackage{graphicx}
\usepackage{amsmath}
\usepackage{amsfonts}
\usepackage{amssymb}
\usepackage{mathrsfs}
\usepackage{bm}

\newcommand{\be}{\begin{eqnarray}}
\newcommand{\ee}{\end{eqnarray}}
\newcommand{\eps}{\epsilon}
\newcommand{\ba}{{\bf a}}
\newcommand{\br}{{\bf r}}
\newcommand{\bx}{{\bf x}}
\newcommand{\by}{{\bf y}}
\newcommand{\bk}{{\bf k}}

\newcommand{\bj}{{\bf j}}
\newcommand{\bp}{{\bf p}}

\newcommand{\bq}{{\bf q}}
\newcommand{\bA}{{\bf A}}

\newcommand{\bH}{{\bf H}}

\newcommand{\bR}{{\bf R}}

\DeclareMathAlphabet{\mathcalligra}{T1}{calligra}{m}{n}
\DeclareMathAlphabet{\mathpzc}{OT1}{pzc}{m}{it} 
\begin{document}
\title{Intrinsic thermal Hall conductivity in the mixed state of d-wave superconductors: from wavepacket dynamics to scaling}
\author{Oskar  Vafek}
\affiliation{National High Magnetic Field Laboratory and Department
of Physics,\\ Florida State University, Tallahasse, Florida 32306,
USA}

\begin{abstract}
Recent numerical calculation of the intrinsic thermal Hall conductivity of nodal d-wave superconductors in the mixed state revealed a rapid increase of this quantity above an onset temperature. Interestingly, this defines a measurable energy scale in an otherwise gapless state. Using the mathematics of magnetic coherent states, in this paper such energy scale is related to a dynamical process associated with the Andreev scattering of an electron wavepacket moving along the constant energy contours in the momentum space. This energy scale is then used to obtain an improved scaling collapse of numerically calculated thermal Hall conductivity in a tight-binding model as a function of temperature, magnetic field and the d-wave pairing amplitude at various band fillings.
The results indicate that the mentioned onset temperature is associated with the ability of the quasiparticle wavepacket to complete its semiclassical orbit before it is appreciably scattered by the superconducting condensate.
\end{abstract}

\date{\today}
\maketitle

\section{Introduction}

The electrical Hall effect is an important technique in materials characterization.
Unfortunately, it provides little useful information below the superconducting transition temperature, even in type II superconductors, for which the magnetic field penetrates the bulk of the sample. This is because no transverse voltage can be established in a superconductor, assuming, as is done thought this paper, that it superconducts i.e. that the vortices are rigidly pinned and not driven into the flux flow regime\cite{parks1969superconductivity}.

On the other hand, a superconducting sample in an external magnetic field $\bH$ and subject to a small heat current density $\bj_Q$, may exhibit a thermal Hall effect, i.e. a temperature gradient perpendicular to both $\bH$ and $\bj_Q$. The thermal Hall conductivity, $\kappa_{xy}$, is then defined as ${j_Q}_x=-\kappa_{xy}\frac{dT}{dy}$.
In the case of extreme type II superconductors considered here, the magnetic field inside the sample is practically uniform, but, because the elementary (Bogoliubov) quasiparticle excitations inside a superconductor are a coherent superposition of an electron and a hole\cite{de1999superconductivity}, the usual theory of thermal Hall effect in normal metals\cite{SmrckaStreda1977} does not apply directly.
Development of such theory is therefore an important step towards extending Hall measurements into the realm of superconductivity.

In a model of non-interacting Bogoliubov quasiparticles, the intrinsic contribution to $\kappa_{xy}$ can be related to the energy dependence of the quasiparticle current Hall response\cite{VafekMelikyanZTPRB2001,CvetkovicVafek2015}. The intrinsic contribution is the part independent of the impurity scattering; it is finite and well defined without any impurities and is expected to dominate in the clean limit. In the superconductor, the quasiparticle current is distinct from the electrical current\cite{DurstLeePRB2000}: if the quasiparticle Hamiltonian operator is $\hat{H}$, the former is proportional to the quasiparticle velocity $\left[\br,\hat{H}\right]/i\hbar$, and the latter to $\partial \hat{H}/\partial \bA$.
Moreover, if the vortices are arranged in a perfect lattice, then Bloch theorem can be employed\cite{FranzTesanovicPRL2000}, and the thermal Hall conductivity can be related to the energy dependence of the Berry curvature of the quasiparticle sub-bands in the (vortex) crystal momentum Brilluoin zone\cite{VafekMelikyanZTPRB2001,CvetkovicVafek2015}.
Using such approach, it was recently shown that at low magnetic field $H$, the intrinsic contribution to $\kappa_{xy}$ exhibits a simple scaling with $H$, and shows a rapid increase from negligible values at low temperature to values of order $1/H$ at a characteristic onset temperature\cite{CvetkovicVafek2015}. In the model used in Ref.\onlinecite{CvetkovicVafek2015}, the onset temperature was shown to increase with increasing $\Delta$, the pairing amplitude of the tight-binding lattice d-wave superconductor whose $H=0$ pairing function is $2\Delta(\cos k_x-\cos k_y)$. While this successfully captures the most important dependence of the onset temperature of $\kappa_{xy}$, the numerical results of the Ref.\onlinecite{CvetkovicVafek2015} at fixed band filling displayed additional (weak) dependence on the Dirac cone anisotropy (see Fig. 2 of Ref.\onlinecite{CvetkovicVafek2015}). This feature has not been explained. In addition, as found in this work, there is an additional dependence of the onset temperature on the band filling in the tight-binding model used in the Ref.\onlinecite{CvetkovicVafek2015}.

As explained in this paper, such features are a consequence of the particular lattice model adopted in Ref.\onlinecite{CvetkovicVafek2015}; the onset temperature dependence on $\Delta$ reported therein indeed captures the main essence of the effect. The mentioned residual dependence can be naturally understood by picturing high energy quasiparticle wavepackets semiclassically moving along the contours of constant (normal) energy. As the energy of the quasiparticle is lowered, the amplitude of its Andreev scattering increases along the anti-nodal portions of its contour, and at some point becomes prohibitively large for the wavepacket to complete its semiclassical orbit before it is appreciably scattered by the superconducting condensate. This marks the energy scale $\eps_*$, which obviously increases with increasing $\Delta$. However, due to the tight-binding dispersion and pairing function used in the model of Ref.\onlinecite{CvetkovicVafek2015}, $\eps_*$ has additional dependence on the Dirac cone anisotropy as well as the band filling. To illustrate the band filling dependence, consider the magnitude of the pairing function on the Fermi surface in the anti-nodal direction, $2\Delta(1-\cos k_F)$; its value depends not only on $\Delta$, but also on $k_F$ which depends on the band filling. If, instead of using $\Delta$ to rescale the temperature, $\eps_*$ is used, then the family of curves for a range of values of the Dirac cone anisotropy and band fillings collapses onto a single scaling curve (see Fig. \ref{fig:kappascaling}). Such improved scaling -- combined with the explicit calculation for the scattering amplitude formulated in continuum and using magnetic coherent states presented below -- therefore strongly supports the above physical picture. It also indicates that $\kappa_{xy}$ may be a way to measure the ability of the quasiparticles to complete their semiclasical orbits before they are appreciably Andreev scattered, providing useful spectroscopic information about unconventional superconductors.

The primary focus of this paper, just as in Ref.\onlinecite{CvetkovicVafek2015}, is the limit $\hbar\omega_c\ll\Delta\ll E_F$, where the Fermi energy $E_F$ is to be measured from the band minimum or maximum, whichever gives the smaller value, and $\omega_c=eH/mc$ is the cyclotron frequency of a point particle with charge $e$ and mass $m$. In this regime a naive perturbation theory in $\Delta$ would appear to break down.
However, as mentioned, the key insight advanced here is that the high energy states must be weakly affected by the pairing term, despite their separation -- set by $\hbar\omega_c$ -- being much smaller than the pairing term amplitude. In order to obtain the energy scale where $\Delta$ ceases acting perturbatively, a first order time dependent perturbation theory calculation is performed using as the starting state a magnetic coherent state\cite{MalkinManko1969}.
Such states are exact eigenstates of the time dependent Schrodinger equation in a uniform magnetic field in the symmetric gauge, but they are not the stationary states -- starting with a stationary state is often assumed in the quantum mechanics textbooks explaining time dependent perturbation theory, but it is, of course, not necessary\cite{BaymBook}. A magnetic coherent state describes a Gaussian wavepacket moving along circular trajectory in the real space with the angular frequency $\omega_c$ and the width of the Gaussian set by the magnetic length $\ell_H=\sqrt{hc/eH}$. In the absence of any other perturbations, the wavepacket width does not change in time.
If one writes the Hamiltonian operator for the electron, $(\bp-\frac{e}{c}\bA)^2/2m$, in terms of the harmonic oscillator ladder operators as $\hbar\omega_c(a^\dagger a+\frac{1}{2})$, and that of the hole, $(\bp+\frac{e}{c}\bA)^2/2m$, as $\hbar\omega_c(b^\dagger b+\frac{1}{2})$, then
the magnetic coherent states $|\alpha\beta \rangle$ are the simultaneous eigenstates of $a$ and $b$.
At finite time the solution of the time dependent Schodinger equation is $|e^{-i\omega_ct}\alpha,\beta \rangle$ when the dynamics is generated by $(\bp-\frac{e}{c}\bA)^2/2m$, and $|\alpha, e^{-i\omega_ct}\beta \rangle$ when by $(\bp+\frac{e}{c}\bA)^2/2m$.

\begin{figure}[t]
\includegraphics[width=0.45\textwidth]{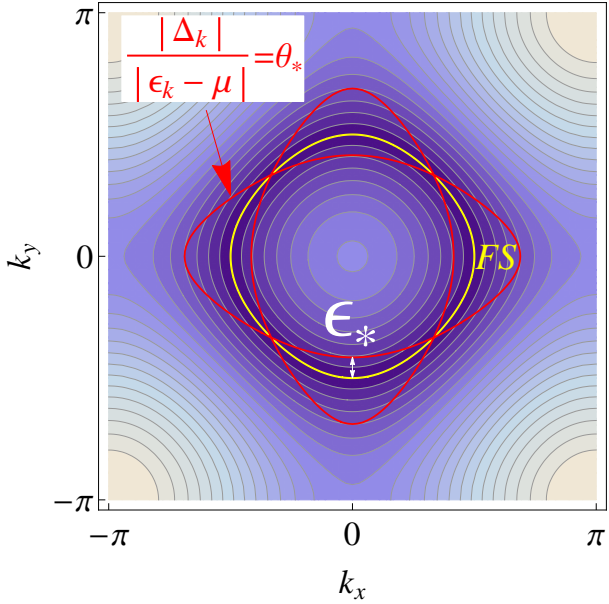}
 \caption{Illustration of the physical process which determines the energy scale, $\eps_*$, for the onset of the intrinsic thermal Hall
 conductivity $\kappa_{xy}$ in the lattice d-wave superconductor. In this figure, representing the 1$^{st}$ Brillouin zone, the pairing amplitude is taken to be $\Delta_{\bk}=2\Delta(\cos k_x-\cos k_y)$, and the normal state dispersion is $\eps_{\bk}=-2\mbox{t}(\cos k_x+\cos k_y)$. The shaded concentric contours are the contours of constant $|\eps_{\bk}-\mu|$, with $\mu=2\mbox{t}$. The Fermi surface (FS), where $|\eps_{\bk}-\mu|=0$ is marked (yellow). The two d-wave shaped lines (red) correspond to contours of constant $\left|\Delta_{\bk}/(\eps_{\bk}-\mu)\right|=\theta_*$. The value of $\theta_*=\frac{1}{\sqrt{2\pi}}\approx 0.4$ chosen here is the same as in Fig. \ref{fig:kappascaling}, where it is shown to result in the collapse of the numerical data.
  \label{fig:dwavescaling}}
\end{figure}

\begin{figure}[t]
\includegraphics[width=0.5\textwidth]{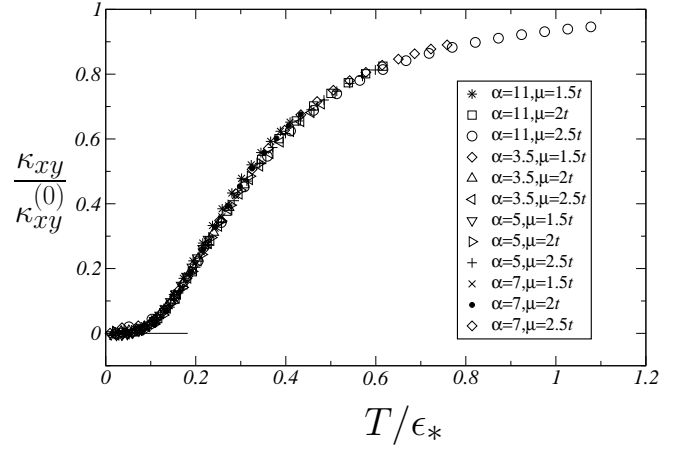}
\caption{Scaling of the thermal Hall conductivity in the mixed state of the lattice d-wave superconductor
 $\kappa_{xy}=\kappa_{xy}(T,\Delta,H,\mu)$, re-scaled by its (free Fermion) normal state value $\kappa_{xy}^{(0)}=\kappa_{xy}(T,0,H,\mu)$. The temperature is rescaled by $\eps_*$, an energy scale associated with Andreev scattering, as discussed in the text. Various values of the Dirac cone dispersion anisotropy $\alpha=v_F/v_{\Delta}=\mbox{t}/\Delta$ and chemical potential $\mu$ are shown in the legend. The magnetic length $\ell_H=28$a in all these calculations, with the square vortex lattice oriented along tight binding unit cell diagonal (see inset of the Fig 2 of Ref.\onlinecite{CvetkovicVafek2015} where magnetic field scaling has been established).
  \label{fig:kappascaling}}
\end{figure}

Thus, the main finding presented in this paper is that, as long as $\hbar\omega_c\ll\Delta\ll E_F$ -- where on the tight-binding lattice with a unit lattice spacing and with the hopping amplitude $\mbox{t}$, $\hbar\omega_c$ should be understood as $4\pi \mbox{t}/\ell_H^2$ -- and, as long as the filling does not coincide with the vicinity of the van Hove singularity, the thermal Hall conductivity has the scaling form
\begin{eqnarray}
\kappa_{xy}(T,\Delta,H,\mu)=\kappa_{xy}(T,0,H,\mu)\times \mathcal{F}\left(\frac{k_BT}{\eps_*}\right),
\end{eqnarray}
where $\kappa_{xy}(T,0,H,\mu)$ is the clean limit normal state thermal Hall conductivity, which obeys free Fermion Wiedemann-Franz law, and scales as $\sim T/H$. Here, $k_B$ is the Boltzman constant, which will be set to unity in what follows, unless stated explicitly otherwise.
The energy scale $\eps_*$ depends on the magnetic field $H$ only through a possible $H$-dependence of $\Delta$ and $\mu$.
$\eps_*$ is to be determined as follows:
consider the normal state dispersion $\eps_{\bk}$ and the pairing function $\Delta_{\bk}$.
Then, as we move along the closed contours of constant $|\eps_\bk-\mu|$ shown in Fig.\ref{fig:dwavescaling}, the quantity $\left|\frac{\Delta_{\bk}}{\eps_\bk-\mu}\right|$ measuring the amplitude of Andreev scattering, varies. For the d-wave superconductor of interest here, this quantity is peaked in the antinodal direction. As we approach the Fermi level, there are two contours of constant $|\eps_\bk-\mu|$, one inside and one outside the Fermi surface, for which the peak value of $\left|\frac{\Delta_{\bk}}{\eps_\bk-\mu}\right|$ is equal to a pure number $\theta_*$ of order unity which will be specified shortly. Then, as shown in Fig. \ref{fig:dwavescaling}, $\eps_*$ is the lesser of the two such values of $|\eps_\bk-\mu|$.

For the specific case of lattice d-wave superconductor considered here, at $H=0$ the pairing amplitude is
$\Delta_{\bk}=2\Delta(\cos k_x-\cos k_y)$ and the normal state dispersion is $\eps_{\bk}=-2\mbox{t}(\cos k_x+\cos k_y)$. This results in $\eps_*=\Delta\frac{4-|\mu/\mbox{t}|}{\theta_*+|\Delta/\mbox{t}|}$. As shown in Fig.\ref{fig:kappascaling}, the scaling collapse of $\kappa_{xy}$ is achieved for $\theta_*\approx 0.4$, a value which is interestingly close to $1/\sqrt{2\pi}$. The resulting scaling function $\mathcal{F}(x)$ is monotonically increasing, approaches $1$ for large $x$, and displays a rapid onset at $x\approx 0.1$ (see Fig. \ref{fig:kappascaling}).
It should also be mentioned that, because we are interested in the limit $\hbar\omega_c\ll \Delta \ll E_F$, the Zeeman effect, corresponding to a trivial shift of all quasiparticle energies, is ignored here.

The rest of the paper provides details of the calculations which lead to the above assertions. In Section II, the mathematics of the magnetic coherent states in symmetric gauge is reviewed. The methods for constructing the pairing order parameter in the vortex state, and in the symmetric gauge, are reviewed in Sec IIa. The time dependent perturbation theory to first order in the pairing term, and in the basis of the magnetic coherent states, is described in Sec IIb. The entire formulation in Sec II is in continuum. The formulation on the discrete tight-binding lattice, along with the formula used to numerically compute the thermal Hall conductivity from numerically diagonalizing the tight-binding Hamiltonian, are reviewed in Sec III. Discussion is in Sec IV, and the details of the perturbative calculation with the magnetic coherent states are delegated to the Appendix.

\section{Magnetic coherent states}
This section follows the original article by Malkin and Man'ko\cite{MalkinManko1969}.
It is included here in order to establish notation and the main mathematical identities which will be used in later sections.
This formulation is in continuum.

In the symmetric gauge $\bA=\frac{1}{2}H\hat{z}\times\br=\frac{1}{2}H\left(-y,x,0\right)$.
The cyclotron frequency is $\omega_c=eH/(mc)$, and let
 $\ell=\sqrt{\hbar c/eH}$. Note that this differs by a factor of $1/\sqrt{2\pi}$
 from the definition of the magnetic length $\ell_H=\sqrt{hc/eH}$ used earlier.

The Schrodinger Hamiltonian operator for the electron in the magnetic field is then
\begin{eqnarray}
\mathcal{H} &=& \frac{\left(\bp-\frac{e}{c}\bA\right)^2}{2m}.
\end{eqnarray}
Let us define a dimensionless variable
\begin{eqnarray}
\xi=(x+iy)/(2\ell),
\end{eqnarray}
and raising a and lowering operators satisfying $\left[a,a^\dagger\right]=1$, where
\begin{eqnarray}
a&=&-\frac{i}{\sqrt{2}}\left(\xi+\frac{\partial}{\partial \xi^*}\right),\\
a^{\dagger}&=&\frac{i}{\sqrt{2}}\left(\xi^*-\frac{\partial}{\partial \xi}\right).
\end{eqnarray}
Then,
\begin{eqnarray}
\mathcal{H}=\hbar\omega_c\left(a^{\dagger}a+\frac{1}{2}\right).
\end{eqnarray}
Note that there is another set of raising and lowering operators, satisfying $\left[b,b^\dagger\right]=1$, where
\begin{eqnarray}
b&=&\frac{1}{\sqrt{2}}\left(\xi^*+\frac{\partial}{\partial \xi}\right),\\
b^{\dagger}&=&\frac{1}{\sqrt{2}}\left(\xi-\frac{\partial}{\partial \xi^*}\right).
\end{eqnarray}
These do not appear explicitly in the Hamiltonian, but, importantly, they commute with the previous ones:
\begin{eqnarray}
\left[a,b\right]&=&\left[a,b^\dagger\right]=0.
\end{eqnarray}
They therefore represent a constant of motion.
For a particle with opposite charge, the Schrodinger Hamiltonian operator can be written as
\begin{eqnarray}
\frac{\left(\bp+\frac{e}{c}\bA\right)^2}{2m}=\hbar\omega_c\left(b^{\dagger}b+\frac{1}{2}\right).
\end{eqnarray}

The "vacuum" state $|00\rangle$ is simultaneously annihilated by $a$ and $b$, and in coordinate representation is given by
\begin{eqnarray}
\langle\br|00\rangle=\frac{1}{\sqrt{2\pi}\ell}e^{-\xi^*\xi}.\end{eqnarray}
This state is used to build coherent states\cite{Glauber1963}.
In order to do so, define the unitary operators
\begin{eqnarray}
\hat{D}(\alpha)&=&e^{\alpha a^{\dagger}-\alpha^* a},\\
\hat{D}(\beta)&=&e^{\beta b^{\dagger}-\beta^* b},
\end{eqnarray}
where $\alpha$ and $\beta$ are two complex c-numbers.
Clearly, the two operators commute:
\begin{eqnarray}
\left[\hat{D}(\alpha),\hat{D}(\beta)\right]&=&0.
\end{eqnarray}
The common coherent state of $a$ and $b$ is
\begin{eqnarray}
|\alpha\beta\rangle &=& \hat{D}(\alpha)\hat{D}(\beta)|00\rangle.
\end{eqnarray}
In the coordinate representation, such state has the form
\begin{eqnarray}
\langle\br|\alpha\beta\rangle &=& \frac{1}{\sqrt{2\pi}\ell}e^{-\xi^*\xi}e^{\sqrt{2}\beta\xi+i\sqrt{2}\alpha\xi^*}
e^{-i\alpha\beta}e^{-\frac{1}{2}\left(|\alpha|^2+|\beta|^2\right)}.\nonumber\\
\end{eqnarray}
This is a Gaussian centered at $\bar{\xi}=\left(\beta^*+i\alpha\right)/\sqrt{2}$ and modulated by the phase which grows linearly with $x$ and $y$.
The kinetic momentum for the electron, $p_x-\frac{e}{c}A_x+i\left(p_y-\frac{e}{c}A_y\right)=\sqrt{2}\hbar a/\ell$, in such a state is peaked at $\sqrt{2}\hbar \alpha/\ell$.
The kinetic momentum for the hole, $p_x+\frac{e}{c}A_x+i\left(p_y+\frac{e}{c}A_y\right)=\sqrt{2}i\hbar b^{\dagger}/\ell$, in such a state is peaked at $\sqrt{2}i\hbar \beta^*/\ell$.

The coherent states form an overcomplete set, and can be used to construct the resolution of identity
\begin{eqnarray}\label{eq:resolution of identity}
\int\frac{d\alpha^*d\alpha}{2\pi i}\int\frac{d\beta^*d\beta}{2\pi i}
\langle \br |\alpha\beta\rangle\langle \alpha\beta|\br'\rangle=\delta(\br-\br').
\end{eqnarray}
Here $\int\frac{d\alpha^*_1d\alpha_1}{2\pi i}(\ldots)=\int_{-\infty}^{\infty}\int_{-\infty}^{\infty}\frac{d\Re e\alpha\;\; d\Im m\alpha}{\pi}(\ldots)$.

\subsection{Pairing order parameter in symmetric gauge}
The development in this section follows the work of T. Kita\cite{Kita1998}.
Because we are dealing with charge $2e$ order parameter, let $\ell_*=\sqrt{\hbar c/2eH}=\ell/\sqrt{2}$.
Then, consider a set of 2D lattice points
\begin{eqnarray}
\bR=n_1\ba_1+n_2\ba_2,
\end{eqnarray}
where $n_1$ and $n_2$ are integers.
The primitive lattice vectors are $\ba_1=(a_{1x},a_{1y},0)$ and $\ba_2=(0,a_2,0)$, where $a_{1x}a_2=2\pi\ell_*^2$.

In the symmetric gauge, the operator which corresponds to the translation by a lattice vector $\bR$, followed by a gauge transformation,
is
\begin{eqnarray}
\hat{T}(\bR)=e^{-i\left(R_yx-R_xy\right)/(2\ell_*^2)}e^{-\bR\cdot\nabla}.
\end{eqnarray}
Note that the exponents commute. This operator commutes with $\left(\bp-\frac{2e}{c}\bA\right)^2$, whose
ground state wavefunction
\begin{eqnarray}
\frac{1}{\sqrt{2\pi}\ell_*}e^{-\left(x^2+y^2\right)/(4\ell^2_*)}
\end{eqnarray}
serves to generate the order parameter; more precisely and as discussed below, its center-of-mass coordinate dependence.

The irreducible representation for the magnetic translation group (see e.g. Ref. \onlinecite{Kita1998}) are
\begin{eqnarray}
D^{(\bq)}(\bR)=e^{-i\bq\cdot\bR-i\pi n_1n_2}.
\end{eqnarray}
Then, at $\bq=0$, the s-wave Abrikosov order parameter can be written as
\begin{eqnarray}\label{eq:opLLL}
&&\Delta\sum_{\bR}e^{i\pi n_1n_2}\hat{T}(\bR)e^{-\frac{1}{4\ell^2_*}\left(x^2+y^2\right)}=\nonumber\\
&&\Delta\sum_{n_1=-\infty}^{\infty}\sum_{n_2=-\infty}^{\infty}
e^{i\pi n_1n_2}e^{-\frac{1}{2}\zeta_{\bR}^*\zeta_{\bR}}e^{2\zeta^*_{\bR}\xi}e^{-2\xi^*\xi}\end{eqnarray}
where
\begin{eqnarray}
\zeta_{\bR}=\frac{R_x+iR_y}{\sqrt{2}\ell_*}.
\end{eqnarray}
The results of this section will be used below to construct the center-of-mass dependence of the pairing amplitude in symmetric gauge.

\subsection{Andreev wavepacket scattering}
The dynamics of the problem we are interested in is generated by the Bogoliubov-de Gennes
Hamiltonian operator
\begin{eqnarray}
\mathcal{H}_{BdG}&=&\left(\begin{array}{cc}
\frac{\left(\bp-\frac{e}{c}\bA\right)^2}{2m}-E_F & \hat\Delta\\
\hat\Delta^\dagger & -\frac{\left(\bp+\frac{e}{c}\bA\right)^2}{2m}+E_F
\end{array}
\right),
\end{eqnarray}
where the center-of-mass coordinate dependence and the relative coordinate of the pairing operator\cite{SimonLeePRL1997} can be expanded as
\begin{eqnarray}
\langle \br' |\hat\Delta |\br\rangle &=&\sum_{j}\Delta_j \Psi_j\left(\frac{\br+\br'}{2}\right)\chi_j(\br-\br')\\
&=&\sum_j\Delta_j \Psi_j\left(\frac{\br+\br'}{2}\right)\int\frac{d^2\bk}{(2\pi)^2}d_j(\bk) e^{i\bk\cdot(\br-\br')}.\nonumber\\
\label{eq:delta matrix element}\end{eqnarray}
In what follows, the restriction in the sum over $j$ will be made to the lowest term and, making use of Eq.(\ref{eq:opLLL}),
\begin{eqnarray}
\Delta_0\Psi_0(\br)=\Delta\sum_{\bR}e^{i\pi n_1n_2}e^{-\frac{1}{2}\zeta_{\bR}^*\zeta_{\bR}}e^{2\zeta^*_{\bR}\xi}e^{-2\xi^*\xi}.
\end{eqnarray}
For nodal d-wave superconductor
\begin{eqnarray}
d_0(\bk)=\frac{k^2_x-k^2_y}{\bk^2}.
\end{eqnarray}
For an s-wave superconductor the above quantity would be equal to unity.

We are now in the position to define our scattering problem.
The Nambu spinor $|\psi_t\rangle$ evolves in time according to
\begin{eqnarray}
i\hbar\frac{\partial}{\partial t}|\psi_t\rangle&=&\mathcal{H}_{BdG}|\psi_t\rangle.
\end{eqnarray}
We write
\begin{eqnarray}
\mathcal{H}_{BdG}=H_0+V,
\end{eqnarray}
where
\begin{eqnarray}
H_0&=&\left(\begin{array}{cc}
\hbar\omega_c\left(a^{\dagger}a+\frac{1}{2}\right)-E_F & 0\\
0 & -\hbar\omega_c\left(b^{\dagger}b+\frac{1}{2}\right)+E_F
\end{array}
\right),\\
V&=&\left(\begin{array}{cc}
0 & \hat\Delta\\
\hat\Delta^\dagger & 0
\end{array}
\right),
\end{eqnarray}
and separate the time evolution due to $H_0$ as
\begin{eqnarray}
|\psi_t\rangle&=&e^{-\frac{i}{\hbar}H_0t}|\psi(t)\rangle.
\end{eqnarray}
Standard time dependent perturbation theory\cite{BaymBook} gives
\begin{eqnarray}
|\psi(t)\rangle&=&|\psi(0)\rangle+\frac{1}{i\hbar}\int_0^{t}dt'e^{\frac{i}{\hbar}H_0t'}Ve^{-\frac{i}{\hbar}H_0t'}|\psi(t')\rangle\\
&\approx&|\psi(0)\rangle+\frac{1}{i\hbar}\int_0^{t}dt'e^{\frac{i}{\hbar}H_0t'}Ve^{-\frac{i}{\hbar}H_0t'}|\psi(0)\rangle +\ldots\nonumber\\
\end{eqnarray}
We are interested in finding $\langle \br|\psi_t\rangle$ given the initial state being the magnetic coherent state, which, without loss of generality we choose to be purely hole-like
\begin{eqnarray}
|\psi(0)\rangle=\left(\begin{array}{c} 0 \\|\alpha_0\beta_0\rangle  \end{array}\right).
\end{eqnarray}
This state is not an eigenstate of $H_0$, but its time evolution due to $H_0$ is known exactly.
At $t=0$, it corresponds to a Gaussian wavepacket peaked at $(x_0+iy_0)/2\ell=(\beta_0^*+i\alpha_0)/\sqrt{2}$.
The time evolution due to $H_0$-only makes the complex variable $\alpha_0$ time independent, and $\beta_0(t)=e^{i\omega_ct}\beta_0$.
Therefore, $\alpha_0$ determines the position of the center of the circle, and $\beta_0(t)$ the radius of, and the angle along, the circle describing the classical motion of the wavepacket. The shape of the wavepacket does not change in time.

We expect that when the initial wavepacket is prepared at an energy far away from the Fermi level, the effect of the pairing term is small, and that such wavepacket remains hole-like and that it continues to move along the circular trajectory.
Therefore, if $|\hbar\omega_c \beta^*_0\beta_0-E_F|\gg \Delta$, there should be no appreciable Andreev scattering.
The goal is to determine the condition on $\alpha_0$ and $\beta_0$ which would mark the transition from the regime where the wavepacket is unaffected by the condensate to the regime where the scattering is significant. Once such condition is identified, it will be utilized to define the energy scale $\eps_*$, which is in turn used to achieve the scaling collapse of the non-perturbative numerical calculation for lattice d-wave superconductor in the mixed state.
Although we are interested in the limit $\hbar\omega_c\ll\Delta\ll E_F$, this limit will be taken only at the end of the perturbative calculation.

\begin{widetext}
We imagine evolving the wavepacket from time $0$ to time $t$, which is of order $\hbar/\Delta$.
Using the resolution of identity in terms of the magnetic coherent states (\ref{eq:resolution of identity})
we find
\begin{eqnarray}\label{eq:tdpt}
\langle \br|\psi_t\rangle
&\approx&
e^{i\left(\frac{1}{2}\omega_c-E_F/\hbar\right)t}
\left[\left(\begin{array}{c} 0 \\ \langle\br |\alpha_0, e^{i\omega_ct}\beta_0\rangle  \end{array}\right)\right.\nonumber\\
&+&\left.\frac{1}{i\hbar}\int_0^{t}dt'
e^{i\left(\omega_c-2E_F/\hbar\right)\left(t'-t\right)}
\int\frac{d\alpha^*_1d\alpha_1}{2\pi i}\int\frac{d\beta^*_1d\beta_1}{2\pi i}
\left(\begin{array}{c} \langle \br
|\alpha_1 e^{-i\omega_c(t-t')},\beta_1\rangle \langle \alpha_1,\beta_1|
\hat{\Delta}|\alpha_0, e^{i\omega_ct'}\beta_0\rangle \\ 0  \end{array}\right)\right].
\end{eqnarray}
The integral over $\alpha_1$ and $\beta_1$ can be performed exactly, and so can the integral over $\bk$, the Fourier wavevector used to define $\chi_0(\br)$ in Eq. (\ref{eq:delta matrix element}).
\end{widetext}
In order to perform the time integral in Eq.(\ref{eq:tdpt}), we now take the limit
\begin{equation}
\frac{\hbar\omega_c}{\Delta}\ll 1\ll \frac{E_F}{\Delta},
\end{equation}
and assume that $t\lesssim \hbar/\Delta$.
\begin{widetext}
After a somewhat lengthy but straightforward calculation (see Appendix for details), one finds that the dominant term for scattered part takes the form
\begin{eqnarray}
&&\frac{1}{i\hbar}\int_0^{t}dt'
e^{i\left(\omega_c-2E_F/\hbar\right)\left(t'-t\right)}
\int\frac{d\alpha^*_1d\alpha_1}{2\pi i}\int\frac{d\beta^*_1d\beta_1}{2\pi i}
\langle \br
|\alpha_1 e^{-i\omega_c(t-t')},\beta_1\rangle \langle \alpha_1,\beta_1|
\hat{\Delta}|\alpha_0, e^{i\omega_ct'}\beta_0\rangle\nonumber\\
&\approx&-\langle\br|\alpha_0\beta_0(t)\rangle
\left(\sum_{\bR}e^{i\pi n_1n_2}e^{-\frac{1}{2}\left|\zeta_{\bR}-2\xi\right|^2}e^{\zeta^*_\bR\xi-\zeta_\bR\xi^*}\right)\nonumber\\
&\times&
\frac{1}{2i\hbar}\left(\frac{\beta_0(t)}{\beta_0^*(t)}+\frac{\beta^*_0(t)}{\beta_0(t)}\right)
\frac{\Delta}{-2i\frac{E_F}{\hbar}+2i\omega_c\beta^*_0(t)\beta_0(t)}
\left(1-e^{\left(2i\frac{E_F}{\hbar}-2i\omega_c\beta^*_0(t)\beta_0(t)\right)t}\right)
\label{eq:dwave wavepacket scattering}
\end{eqnarray}
where $\beta_0(t)=\beta_0e^{i\omega_ct}$.
\end{widetext}
Thus, in the stated limit, the wavepacket is appreciably
Andreev scattered only if
\begin{eqnarray}
\frac{\Delta\left|\frac{\beta_0(t)}{\beta_0^*(t)}+\frac{\beta^*_0(t)}{\beta_0(t)}\right|}{
2\left|E_F-\hbar\omega_c\beta^*_0(t)\beta_0(t)\right|}\gtrsim 1.
\end{eqnarray}
Recall that the typical value of the kinetic momentum operator for the hole, $p_x+\frac{e}{c}A_x+i\left(p_y+\frac{e}{c}A_y\right)$, is $\sqrt{2}i\hbar\beta^*/\ell$. Therefore, we can interpret the term in the numerator in the above expression as
the d-wave form factor amplitude. Such term is of course peaked in the anti-nodal direction. The term in the denominator represents the difference between the typical energy of the hole wavepacket, i.e. the peak value of $\left(\bp+\frac{e}{c}\bA\right)^2/2m$, and the Fermi energy.
The Andreev scattering is therefore maximized when the wavepacket is near the Fermi surface in the anti-nodal direction.
If it is far from the Fermi surface, or is near the node, the wavepacket continues moving along the constant energy contours at $\Delta=0$,
essentially as if the system was a normal metal. The above condition therefore marks the transition from the energy regime where the wavepacket continues to move according to the semiclassical dynamics along the contours of constant energy, essentially undisturbed by the superconducting condensate, and the lower energy regime where the Andreev scattering occurs on time scales much shorter than $1/\omega_c$ i.e. the time scale the wavepacket would need to complete the orbit. It is in the lower energy regime that we find the suppression of the thermal Hall conductivity, as discussed in the next section.

\section{Lattice d-wave numerical calculation and scaling}
\subsection{Tight-binding model}
In this section we resort to the numerical calculation of the intrinsic contribution to $\kappa_{xy}$ along the lines discussed in Refs. (\onlinecite{VafekMelikyanZTPRB2001}) and (\onlinecite{CvetkovicVafek2015}).
\begin{widetext}
We work on a two dimensional square lattice of spacing $a$ -- that we set to unity -- and perpendicular magnetic field $\bH$.
The tight-binding Hamiltonian describing the excitations is
\begin{eqnarray}\label{eq:hamiltonian tight binding}
\mathcal{H}=\sum_{\br}\left(\left(\sum_{\bm{\delta}=\hat{\bx},\hat{\by}}t_{\br,\br+\bm{\delta}}c^{\dagger}_{\br,\sigma}c_{\br+\bm{\delta},\sigma}
+\Delta_{\br,\br+\bm{\delta}}\left(c^{\dagger}_{\br,\uparrow}c^{\dagger}_{\br+\bm{\delta},\downarrow}-c^{\dagger}_{\br,\downarrow}c^{\dagger}_{\br+\bm{\delta},\uparrow}\right)+H.c.\right)-\mu c^{\dagger}_{\br,\sigma}c_{\br,\sigma}
\right).
\end{eqnarray}
Here $c_{\br,\sigma}$ is the electron annihilation operator on the tight-binding lattice site $\br$, not to be confused with the vortex lattice.
\end{widetext}
The sum over the spin projection $\sigma=\uparrow$ or $\downarrow$ in the first and the last term of Eq.(\ref{eq:hamiltonian tight binding}) is implicit; $H.c.$ stands for Hermitian conjugation.
The (nearest neightbor) hopping occurs in the presence of the uniform magnetic field, encoded in the Peierls phase factor
\begin{eqnarray}
t_{\br,\br+\bm{\delta}}=-\mbox{t}e^{-iA_{\br,\br+\bm{\delta}}}.
\end{eqnarray}
The magnetic flux $\Phi$ through the elementary tight-biding plaquette appears through the link integral of the vector potential
\begin{eqnarray}
A_{\br,\br+\hat{\bx}}&=&-\pi y\frac{\Phi}{\phi_0},\\
A_{\br,\br+\hat{\by}}&=&\pi x\frac{\Phi}{\phi_0}.
\end{eqnarray}
The electronic flux quantum is $\phi_0=hc/e$.

The ansatz for the tight-binding lattice pairing term is
\begin{eqnarray}\label{eq:lattice dwave vortex amplitude}
\Delta_{\br,\br+\bm{\delta}}&=&\Delta_{\bm{\delta}}e^{i\theta(\br)}e^{\frac{i}{2}\int_{\br}^{\br+\bm{\delta}}d{\bf l}\cdot\nabla \theta}\\
\Delta_{\hat{\bx}}&=&-\Delta_{\hat{\by}}=\Delta,
\end{eqnarray}
and the line integral is over the nearest neighbor link. Vortex positions, $\bR_j$, are inside the centers of some of the elementary plaquettes.
They enter the pairing term through $\theta(\br)$ which is chosen to be the solution of the continuum London's equations
\begin{eqnarray}
\nabla\times\nabla \theta(\br)&=&2\pi\hat{{\bf z}}\sum_{j}\delta(\br-\bR_j)\\
\nabla\cdot\nabla \theta(\br)&=&0.
\end{eqnarray}
Vortices are positioned in the square lattice arrangement, with the vortex lattice at $45^{\circ}$ relative to the underlying tight-binding lattice.
As shown in Ref.\onlinecite{CvetkovicVafek2015} the results discussed below are largely independent of this choice.
Each $L\times L$ magnetic unit cell is threaded by magnetic flux $hc/e$ and contains a pair of vortices. Although the notation in this section uses the upper case letter $L$ to denote the period of the magnetic unit cell, because the square vortex lattice is considered, in the tight-binding lattice units, it is equivalent to $\ell_H$ introduced earlier.

The closed form solution of the London's equations for the pairing field with such arrangement of vortices\cite{CvetkovicVafek2015}, ensuring that the superfluid velocity,
which is proportional to $\frac{\hbar}{2}\nabla\theta(\br)-\frac{e}{c}\bA(\br)$, vanishes on average, is
\begin{eqnarray}
\theta(\br)=\sum_{j=1}^{2}\left(\arg\left[\sigma(z-z_j;\omega,\omega')\right]+\frac{\pi}{2iL^2}(zz^*_j-z^*z_j)\right).\nonumber\\
\end{eqnarray}
Here, $z=x+iy$ (in tight-binding lattice units), $z_j$'s denote the vortex positions inside the magnetic unit cell, and $\sigma(z;\omega,\omega')$ is the Weierstrass $\sigma$ functions with periods $\omega=L$ and $\omega'=iL$.

The singular gauge transformation\cite{FranzTesanovicPRL2000,VafekMelikyanFranzZTPRB2001,VafekMelikyanPRL2006} turns the hopping and the pairing terms in Eq.(\ref{eq:hamiltonian tight binding}) periodic with the periodicity $L\times L$, enabling the use of Bloch theorem.
Performing the operator change of variables
\begin{eqnarray}
\left(\begin{array}{c}
c_{\br,\uparrow} \\ c^{\dagger}_{\br,\downarrow}
\end{array}\right)=\frac{1}{\sqrt{N_{uc}}}\sum_{\bk}
\left(\begin{array}{c}
e^{\frac{i}{2}\theta(\br)}\psi_{\br,\uparrow}(\bk) \\ e^{-\frac{i}{2}\theta(\br)}\psi_{\br,\downarrow}(\bk)
\end{array}\right)\end{eqnarray}
where $N_{uc}$ is the number of magnetic unit cells in the entire lattice, $\psi_{\br,\sigma}(\bk)$ is periodic in $\br$ with the periodicity of the magnetic unit cell, and $\bk$ is within the magnetic Brilluoin zone
$-\frac{\pi}{L}\leq k_{x,y} \leq \frac{\pi}{L}$.

The factors $e^{\frac{i}{2}\theta(\br)}$ must be handled with care due to the sign ambiguity associated with taking the square-root of a complex number.
To start with, we connect vortices pairwise within each magnetic unit cell with branch-cuts, which are themselves periodic with the periodicity of the magnetic unit cell, and which intersect the elemenary tight-binding links.
We chose the sign of the square-root such that the following identity holds
\begin{eqnarray}\label{eq:branchcut identity}
e^{\frac{i}{2}\theta(\br+\bm{\delta})}e^{-\frac{i}{2}\theta(\br)}&=&z^{(2)}_{\br+\bm{\delta},\br}e^{\frac{i}{2}\int_{\br}^{\br+\bm{\delta}}d{\bf l}\cdot\nabla \theta}.
\end{eqnarray}
In the above, just as in Eq.(\ref{eq:lattice dwave vortex amplitude}), the line integral is again along the nearest neighbor link.
The periodic factor $z^{(2)}_{\br+\bm{\delta},\br}=1$ on each nearest neighbour
link except the ones intersecting the branch cut where $z^{(2)}_{\br+\bm{\delta},\br}=-1$.
The identity between the site factors on the left hand side of the Eq.(\ref{eq:branchcut identity}) and the link factors on the right hand side of (\ref{eq:branchcut identity})
follows from considering products over the links forming closed clockwise
loops around elementary tight-binding plaquettes. The left hand side must give $+1$ around each
such elementary loop, regardless of whether such a loop contains a vortex, because it consists of a product of complex numbers with unit magnitude on
each site.
On the other hand, such a closed loop product formed from $e^{\frac{i}{2}\int_{\br}^{\br+\bm{\delta}}d{\bf l}\cdot\nabla \theta}$ must give $-1$ if the loop contains a vortex and $+1$ if it does not,
because $\oint d{\bf l}\cdot\nabla \theta=\pm 2\pi$ in the first case and $\oint d{\bf l}\cdot\nabla \theta=0$ in the second.
For $L=28$ considered here, inside the first Brilloin zone, the factor $e^{\frac{i}{2}\int_{\br}^{\br+\bm{\delta}}d{\bf l}\cdot\nabla \theta}$ can be conveniently replaced by
$\left(1+e^{i\theta(\br+\bm{\delta})}e^{-i\theta(\br)}\right)/|1+e^{i\theta(\br+\bm{\delta})}e^{-i\theta(\br)}|$.
This way, only site variables enter the numerical calculation, and the link integrals need not be performed.

The Heisenberg equations of motion
\begin{eqnarray}
i\hbar\frac{\partial}{\partial t}\psi_{\br,\sigma}(\bk) &=& \left[\psi_{\br,\sigma}(\bk),\mathcal{H}\right]\nonumber\\
&=& \hat{H}_{BdG}(\bk)\psi_{\br,\sigma}(\bk),
\end{eqnarray}
define the tight-binding lattice Bogoliubov-de Gennes single particle Bloch Hamiltonian operator, $\hat{H}_{BdG}(\bk)$, whose discrete eigenvalues, $E_n(\bk)$, and eigenstates $|n\bk\rangle$,
are labeled by the magnetic sub-band index $n$. For each $\bk$, there are $2L^2$ such eigenstates.

\subsection{Thermal Hall conductivity}

As mentioned at the end of the previous section, we denote by $|n\bk\rangle$ the eigenfunction of $\hat{H}_{BdG}(\bk)$ with energy $E_n(\bk)$
\begin{eqnarray}
\hat{H}_{BdG}(\bk)|n\bk\rangle=E_n(\bk)|n\bk\rangle.
\end{eqnarray}
Then, the thermal Hall conductivity at temperature $T$ has been shown to be given by\cite{VafekMelikyanZTPRB2001,VafekBoulder2014}
\begin{eqnarray}
\kappa_{xy}=\frac{1}{\hbar T}\int_{-\infty}^{\infty}d\xi
\xi^2\left(-\frac{\partial f(\xi)}{\partial\xi}\right)\tilde\sigma_{xy}(\xi)
\end{eqnarray}
where the Fermi occupation factor is
\begin{eqnarray}
f(\xi)=\frac{1}{e^{\xi/(k_BT)}+1},
\end{eqnarray}
and
\begin{widetext}
\begin{eqnarray}\label{eq:TKNN}
\tilde\sigma_{xy}(\xi)
&=&\frac{1}{i}\int\frac{d^2{\bf k}}{(2\pi)^2}\sum_{E_m({\bf k})<\xi<E_n({\bf k})}
\frac{\left\langle m{\bf k} \bigg |\frac{\partial \hat{H}_{BdG}({\bf k})}{\partial k_{x}} \bigg | n{\bf k}\right\rangle
\left\langle n{\bf k} \bigg |\frac{\partial \hat{H}_{BdG}({\bf k})}{\partial k_y} \bigg | m{\bf k}\right\rangle
-(x\leftrightarrow y)
}{\left(E_m({\bf k})-E_n({\bf k})\right)^2}.
\end{eqnarray}
In the above, the double sum over the magnetic sub-band quantum labels $m$ and $n$ is to be performed subject to the stated restriction that for the given $\bk$, $E_m({\bf k})<\xi$ and $E_n({\bf k})>\xi$.
\end{widetext}
It is well known that the above formula can be written as the sum over occupied bands' $\bk$-space integral over the Berry curvature\cite{TKNN1982,Kohmoto1985}:
\begin{eqnarray}
\tilde\sigma_{xy}(\xi)
&=&\frac{1}{2\pi}\sum_{n}\left(\frac{1}{2\pi i}\int_{E_n(\bk)<\xi} d^2\bk \left(\hat{z}\cdot\nabla_{\bk}\times\hat{A}_n(\bk)\right)\right)\nonumber\\
&=&\frac{C}{2\pi},
\end{eqnarray}
where
\begin{eqnarray}
\hat{A}_n(\bk)&=&\langle n\bk|\nabla_{\bk}|n\bk\rangle.
\end{eqnarray}
For each fully occupied band, the integral extends over the entire magnetic Brillouin zone, and the occupied band contribution to $C$ is an integer\cite{TKNN1982,Kohmoto1985,VafekMelikyanZTPRB2001}, the first Chern number.

Therefore, determining the energy dependence of the $\bk$-space integral over the Berry curvature leads to finding the temperature dependence of the intrinsic thermal Hall conductivity.

For the case of lattice d-wave superconductor considered here, (\ref{eq:hamiltonian tight binding}), at $H=0$ the pairing amplitude is
$\Delta_{\bk}=2\Delta(\cos k_x-\cos k_y)$ and the normal state dispersion is $\eps_{\bk}=-2\mbox{t}(\cos k_x+\cos k_y)$.
As discussed in the introduction, the condition
\begin{eqnarray}\label{eq:andreev contour}
\left|\frac{\Delta_{\bk}}{\eps_\bk-\mu}\right|=\theta_*
\end{eqnarray}
results in two solutions (shown by red lines in Fig.\ref{fig:dwavescaling}). Because of the particle-hole asymmetry in the tight-binding dispersion, the value of $|\eps_{\bk}-\mu|$ in the anti-nodal direction along the two contours given by Eq.(\ref{eq:andreev contour}) is not the same.  The lower of the two values of $|\eps_{\bk}-\mu|$ is
\begin{eqnarray}
\eps_*=\Delta\frac{4-|\mu/\mbox{t}|}{\theta_*+|\Delta/\mbox{t}|}.
\end{eqnarray}
The result of the numerical calculation for the model in Eq.(\ref{eq:hamiltonian tight binding}), with the temperature rescaled by $\eps_*$ with $\theta_*=1/\sqrt{2\pi}$ and the $\kappa_{xy}$ with the value for $\Delta$ set to zero, is shown in Fig.\ref{fig:kappascaling}. Because the scaling with magnetic field has already been established, as was the independence on the vortex lattice geometry\cite{CvetkovicVafek2015}, the above was computed for a single value of the magnetic length $L=28$ and square vortex lattice.
The method used here for an efficient computation of $\kappa_{xy}$ has been detailed in Ref.\onlinecite{CvetkovicVafek2015}.

\section{Summary}

The goal of this paper is to provide a physical picture which explains the existence of the onset temperature scale found in numerical calculations of the intrinsic thermal Hall conductivity in the mixed state of the nodal d-wave superconductor. Such picture is based the calculation of the scattering of a magnetic coherent state within time-dependent perturbation theory and identifying an energy scale at which such scattering starts interfering with the ability of a wavepacket to complete its semiclassical orbit. Additionally, the results of the numerical calculation of $\kappa_{xy}$ performed on a tight-binding lattice for the d-wave superconductor in the mixed state are shown to collapse well onto a single scaling curve (Fig.\ref{fig:kappascaling}), provided that the energy scale identified using the mentioned physical picture is used as the unit of temperature.  These results show negligible dependence on the vortex core size as well as on the vortex lattice geometry. Such feature is also manifest within the wavepacket calculation.
Similar calculation was performed in the case of a lattice s-wave superconductor, with on-site pairing term, which, unlike its d-wave counterpart, does not have $\bk$-dependence. In the s-wave case, the $\mu$ dependence of the onset temperature -- which in the d-wave case amounted to $4-|\mu/\mbox{t}|$ -- was absent.

These findings may help establish measurements of $\kappa_{xy}$ in very clean samples as a way to study the momentum structure of the pairing function in magnetic field via the bulk Hall transport method.

\section {Acknowledgments}
This work was supported by the NSF CAREER award under
Grant No. DMR-0955561, NSF Cooperative Agreement No. DMR-0654118, and the State of Florida.

\newpage
\begin{widetext}
\appendix
\section{Details of the Andreev scattering of the magnetic coherent states}
A somewhat lengthy, but otherwise straightforward calculation, leads to
\begin{eqnarray}
&&\int\frac{d\alpha^*_1d\alpha_1}{2\pi i}\int\frac{d\beta^*_1d\beta_1}{2\pi i}
\langle \br
|\alpha_1 e^{-i\omega_c(t-t')},\beta_1\rangle \langle \alpha_1,\beta_1|
\hat{\Delta}|\alpha_0, e^{i\omega_ct'}\beta_0\rangle=\nonumber\\
&=&\frac{1}{\sqrt{2\pi}\ell}e^{-\xi^*\xi}e^{-\frac{i}{\sqrt{2}}e^{i\omega_c(t'-t)}\alpha_0\xi^*}
e^{\frac{i}{2}\alpha_0e^{i\omega_ct'}\beta_0}e^{-\frac{1}{2}\left(|\alpha_0|^2+|\beta_0|^2\right)}
\nonumber\\
&\times&\frac{1}{1+\frac{1}{2}e^{i\omega_c(t'-t)}}
e^{\frac{1}{1+\frac{1}{2}e^{i\omega_c(t'-t)}}\left(\sqrt{2}\xi+e^{i\omega_c(t'-t)}\frac{i}{2}\alpha_0\right)
\left(-\frac{1}{2}e^{i\omega_ct'}\beta_0+\frac{1}{\sqrt{2}}e^{i\omega_c(t'-t)}\xi^*\right)}\nonumber\\
&\times&\frac{\Delta}{2}
\sum_{\bR}e^{i\pi n_1n_2}e^{-\frac{1}{2}\zeta^*_{\bR}\zeta_{\bR}}e^{\zeta^*_\bR\left(\sqrt{2}\frac{\sqrt{2}\xi
+i\alpha_0\left(1+e^{i\omega_c(t'-t)}\right)}{2+e^{i\omega_c(t'-t)}}\right)}
\left(\frac{1}{j^2}+\frac{1}{j'^2}\right)\left(1+\left(\frac{jj'}{\rho}-1\right)e^{\frac{jj'}{\rho}}\right)
\end{eqnarray}
\end{widetext}
where
\begin{eqnarray}
\rho&=&2\left(\frac{1-\frac{1}{2}e^{i\omega_c(t'-t)}}{1+\frac{1}{2}e^{i\omega_c(t'-t)}}\right),\\
j&=&\frac{i}{1+\frac{1}{2}e^{i\omega_c(t'-t)}}\left(\sqrt{2}i\alpha_0-2\xi\right),\\
j'&=&\frac{ie^{i\omega_c(t'-t)}}{1+\frac{1}{2}e^{i\omega_c(t'-t)}}\left(\sqrt{2}e^{i\omega_ct}\beta_0-2\xi^*+\zeta^*_\bR\right).
\end{eqnarray}
To obtain the above, first the overlap $\langle \alpha_1,\beta_1|
\hat{\Delta}|\alpha_0, e^{i\omega_ct'}\beta_0\rangle$ is calculated in terms of the momentum integral; evaluation of
the momentum integral is postponed until the the integrals over $\alpha_1$ and $\beta_1$ are performed. To calculate the integral
over the momentum appearing in the d-wave form factor $d(\bk)$, the denominator of $(k^2_x-k^2_y)/\bk^2$ is rewritten using the identity
$\bk^{-2}=\int_0^{\infty}d\lambda e^{-\lambda\bk^2}$ and the momentum integral, which is a product of a Gaussian and a polynomial -- even when the entire (lengthy) expression is considered -- is performed before the $\lambda$ integral.

The above formula holds generally for any value of the ratio of the cyclotron frequency and the pairing amplitude.
In order to perform the time integral in Eq.(\ref{eq:tdpt}) of the main text, the limit of interest is taken
\begin{equation}
\frac{\hbar\omega_c}{\Delta}\ll 1\ll \frac{E_F}{\Delta}.
\end{equation}
It is also assumed that that the time duration does not exceed the time scale set by the pairing amplitude, i.e. that $t\lesssim 1/\Delta$. The terms containing complicated $t$ dependence in the exponential can now be expanded to linear order in $t$. The resulting $t$-integrals are elementary. We postpone performing them for the sake of clarity, and instead rearrange the terms in order to reveal their physical content.
Judiciously completing the squares, we find that the resulting expression can be brought into the form
\begin{widetext}
\begin{eqnarray}
&&\frac{1}{i\hbar}\int_0^{t}dt'
e^{i\left(\omega_c-2E_F/\hbar\right)\left(t'-t\right)}
\int\frac{d\alpha^*_1d\alpha_1}{2\pi i}\int\frac{d\beta^*_1d\beta_1}{2\pi i}
\langle \br
|\alpha_1 e^{-i\omega_c(t-t')},\beta_1\rangle \langle \alpha_1,\beta_1|
\hat{\Delta}|\alpha_0, e^{i\omega_ct'}\beta_0\rangle\nonumber\\
&\approx&
\frac{2}{3}
\left(\sum_{\bR}e^{i\pi n_1n_2}e^{-\frac{1}{2}\left|\zeta_{\bR}-\frac{2}{3}\left(\xi+i\sqrt{2}\alpha_0\right)\right|^2}
e^{\left(\zeta_\bR^*\frac{1}{3}\left(\xi+i\sqrt{2}\alpha_0\right)-\zeta_\bR\frac{1}{3}\left(\xi^*-i\sqrt{2}\alpha^*_0\right)\right)}\right)
\nonumber\\
&\times&\frac{1}{\sqrt{2\pi}\ell}e^{-\frac{1}{9}\left|\xi-\left(\frac{i}{\sqrt{2}}\alpha_0-\frac{3}{\sqrt{2}}\beta_0^*(t)\right)\right|^2}
e^{\frac{i}{6}\left(\alpha_0\beta_0(t)+\alpha^*_0\beta^*_0(t)\right)}
e^{-\frac{1}{3\sqrt{2}}\left(i\alpha^*_0+\beta_0(t)\right)\xi+\frac{1}{3\sqrt{2}}\left(-i\alpha_0+\beta^*_0(t)\right)\xi^*}
\nonumber\\
&\times&
\frac{\Delta}{2i\hbar}\int_0^{t}dt'
e^{-2i\frac{E_F}{\hbar}\left(t'-t\right)}\left(\frac{1}{j_0^2}+\frac{1}{j_0'^2}\right)\left(
e^{\frac{4}{9}i\omega_c(t'-t)\xi^*\xi}
e^{\frac{2}{9}i\omega_c(t'-t)\left(i\alpha_0\beta_0(t)-i\sqrt{2}\alpha_0\xi^*-\sqrt{2}\xi\beta_0(t)\right)}
e^{-\frac{1}{9}\zeta^*_\bR\left(\sqrt{2}\alpha_0+2i\xi\right)\omega_c(t'-t)}
\right)\nonumber\\
&+&
\frac{2}{3}
\left(\sum_{\bR}e^{i\pi n_1n_2}e^{-\frac{1}{2}\left|\zeta_{\bR}-2\xi\right|^2}e^{\zeta^*_\bR\xi-\zeta_\bR\xi^*}\right)\nonumber\\
&\times&
\frac{1}{\sqrt{2\pi}\ell}
e^{-\left|\xi-\left(\frac{i}{\sqrt{2}}\alpha_0+\frac{1}{\sqrt{2}}\beta_0^*(t)\right)\right|^2}
e^{-\frac{i}{2}\left(\alpha_0\beta_0(t)+\alpha^*_0\beta^*_0(t)\right)}
e^{\frac{1}{\sqrt{2}}\left(i\alpha^*_0+\beta_0(t)\right)\xi+\frac{1}{\sqrt{2}}\left(i\alpha_0-\beta^*_0(t)\right)\xi^*}
\nonumber\\
&\times&
\frac{\Delta}{2i\hbar}\int_0^{t}dt'
e^{-2i\frac{E_F}{\hbar}\left(t'-t\right)}\left(\frac{1}{j_0^2}+\frac{1}{j_0'^2}\right)\left(\frac{j_0j_0'}{\rho_0}-1\right)
e^{-2i\omega_c(t'-t)(2\xi^*\xi+i\alpha_0\beta_0(t)-\sqrt{2}\left(i\alpha_0\xi^*+\xi\beta_0(t)\right))}
e^{\zeta^*_\bR\left(\sqrt{2}\alpha_0+2i\xi\right)\omega_c(t'-t)},
\label{eq:dwave wavepacket scattering appendix}
\end{eqnarray}
\end{widetext}
where $\beta_0(t)=\beta_0e^{i\omega_ct}$ and
\begin{eqnarray}
\frac{1}{j_0^2}&=&-\frac{9}{4}\frac{1}{\left(\sqrt{2}i\alpha_0-2\xi\right)^2},\\
\frac{1}{j_0'^2}&=&-\frac{9}{4}\frac{1}{\left(\sqrt{2}\beta_0e^{i\omega_ct}-2\xi^*+\zeta^*_\bR \right)^2},\\
\frac{jj'}{\rho_0}&=&-\frac{2}{3}\left(\sqrt{2}i\alpha_0-2\xi\right)\left(\sqrt{2}\beta_0e^{i\omega_ct}-2\xi^*+\zeta^*_\bR\right).
\end{eqnarray}
Eq.(\ref{eq:dwave wavepacket scattering appendix}) has the form of a sum of two terms, representing the superposition of wavepackets.

To analyze the first term in the Eq.(\ref{eq:dwave wavepacket scattering appendix}),
note that the sum over $\bR$ in the parenthesis corresponds to the superposition of Gaussians in $\xi$, whose centers are determined by the value of
$\zeta_R$. The Gaussians are modulated by a pure phase factor. Therefore, if the value of $\xi$ is held fixed, then there is a value of $\bR$ for which $\zeta_{\bR}$ comes close to maximizing the magnitude of the Gaussian. The next term multiplying the sum over $\bR$ in the parenthesis is also a Gaussian in $\xi$ multiplied by a pure phase. It is peaked at
\begin{eqnarray}
&&\xi_{peak}=\frac{i}{\sqrt{2}}\alpha_0-\frac{3}{\sqrt{2}}\beta^*_0(t).
\end{eqnarray}
In the subsequent time integral, the values of $\xi$ and $\xi^*$ in the exponential are multiplied by a power of $\omega_c$. Because the time interval is restricted to $t\lesssim \hbar/\Delta$, the values of $\xi$ and $\zeta_{\bR}$ may be replaced by their peak values inside the time integral
\begin{eqnarray}
&&\zeta^{peak}_\bR\approx \frac{2}{3}\left(\xi_{peak}+i\sqrt{2}\alpha_0\right)
=\sqrt{2}\left(i\alpha_0-\beta_0^*(t)\right)\nonumber\\
\end{eqnarray}
Similarly,
\begin{eqnarray}
&&\frac{1}{j_0^2}+\frac{1}{j_0'^2}\approx-\frac{1}{8}\left(\frac{1}{{\beta_0^*}^2(t)}+\frac{1}{{\beta_0}^2(t)}\right)\\
&&e^{\frac{4}{9}i\omega_c(t'-t)\xi^*\xi}
e^{\frac{2}{9}i\omega_c(t'-t)\left(i\alpha_0\beta_0(t)-i\sqrt{2}\alpha_0\xi^*-\sqrt{2}\xi\beta_0(t)\right)}\nonumber\\
&\times&e^{-\frac{1}{9}\zeta^*_\bR\left(\sqrt{2}\alpha_0+2i\xi\right)\omega_c(t'-t)}
\approx
e^{i\omega_c(t'-t)2\beta^*_0(t)\beta_0(t)}
\end{eqnarray}
When the time integral is performed, the denominator containing $E_F-\hbar\omega_c\beta_0^*\beta_0$ appears. In the state limit, this forces the entire expression to vanish, unless the value of $\beta_0$ for the wavepacket of interest is such that $\hbar\omega_c\beta_0^*\beta_0\approx E_F$.
However, because of the term $j_0^{-2}+j_0'^{-2}$, such scattered wavepacket is effectively suppressed by one power of $\hbar\omega_c/E_F$.

The second term Eq.(\ref{eq:dwave wavepacket scattering appendix}) is peaked at
\begin{eqnarray}
\xi_{peak}&=&\frac{i}{\sqrt{2}}\alpha_0+\frac{1}{\sqrt{2}}\beta^*_0(t)
\end{eqnarray}
which makes the sum over $\bR$ dominated by the value
\begin{eqnarray}
\zeta^{peak}_\bR&\approx& 2\xi_{peak}=\sqrt{2}\left(i\alpha_0+\beta_0^*(t)\right),
\end{eqnarray}
allowing for the replacements
\begin{eqnarray}
\frac{1}{j_0^2}+\frac{1}{j_0'^2}&\approx&-\frac{9}{8}\left(\frac{1}{{\beta_0^*}^2(t)}+\frac{1}{{\beta_0}^2(t)}\right),\\
\frac{j_0j'_0}{\rho_0}&\approx&\frac{4}{3}\beta_0^*(t)\beta_0(t),
\end{eqnarray}
and
\begin{eqnarray}
&&e^{-2i\omega_c(t'-t)(2\xi^*\xi+i\alpha_0\beta_0(t)-\sqrt{2}\left(i\alpha_0\xi^*+\xi\beta_0(t)\right))}\nonumber\\
&\times& e^{\zeta^*_\bR\left(\sqrt{2}\alpha_0+2i\xi\right)\omega_c(t'-t)}
\approx e^{i\omega_c(t'-t)2\beta^*_0(t)\beta_0(t)}
\end{eqnarray}
Because the factor $j_0j'_0/\rho_0$ contains an additional factor of $|\beta_0|^2$ in the numerator, the suppression appearing in the first term discussed above is absent.
Therefore, the dominant term is
\begin{widetext}
\begin{eqnarray}
&&\frac{1}{i\hbar}\int_0^{t}dt'
e^{i\left(\omega_c-2E_F/\hbar\right)\left(t'-t\right)}
\int\frac{d\alpha^*_1d\alpha_1}{2\pi i}\int\frac{d\beta^*_1d\beta_1}{2\pi i}
\langle \br
|\alpha_1 e^{-i\omega_c(t-t')},\beta_1\rangle \langle \alpha_1,\beta_1|
\hat{\Delta}|\alpha_0, e^{i\omega_ct'}\beta_0\rangle\nonumber\\
&\approx&-
\left(\sum_{\bR}e^{i\pi n_1n_2}e^{-\frac{1}{2}\left|\zeta_{\bR}-2\xi\right|^2}e^{\zeta^*_\bR\xi-\zeta_\bR\xi^*}\right)
\frac{1}{\sqrt{2\pi}\ell}
e^{-\left|\xi-\left(\frac{i}{\sqrt{2}}\alpha_0+\frac{1}{\sqrt{2}}\beta_0^*(t)\right)\right|^2}
e^{-\frac{i}{2}\left(\alpha_0\beta_0(t)+\alpha^*_0\beta^*_0(t)\right)}
e^{\frac{1}{\sqrt{2}}\left(i\alpha^*_0+\beta_0(t)\right)\xi+\frac{1}{\sqrt{2}}\left(i\alpha_0-\beta^*_0(t)\right)\xi^*}
\nonumber\\
&\times&
\frac{\Delta}{2i\hbar}\left(\frac{\beta_0(t)}{\beta_0^*(t)}+\frac{\beta^*_0(t)}{\beta_0(t)}\right)\int_0^{t}dt'
e^{-2i\frac{E_F}{\hbar}\left(t'-t\right)}
e^{2i\omega_c(t'-t)\beta^*_0(t)\beta_0(t)}
\nonumber\\
&=&-
\left(\sum_{\bR}e^{i\pi n_1n_2}e^{-\frac{1}{2}\left|\zeta_{\bR}-2\xi\right|^2}e^{\zeta^*_\bR\xi-\zeta_\bR\xi^*}\right)
\frac{1}{\sqrt{2\pi}\ell}
e^{-\left|\xi-\left(\frac{i}{\sqrt{2}}\alpha_0+\frac{1}{\sqrt{2}}\beta_0^*(t)\right)\right|^2}
e^{-\frac{i}{2}\left(\alpha_0\beta_0(t)+\alpha^*_0\beta^*_0(t)\right)}
e^{\frac{1}{\sqrt{2}}\left(i\alpha^*_0+\beta_0(t)\right)\xi+\frac{1}{\sqrt{2}}\left(i\alpha_0-\beta^*_0(t)\right)\xi^*}
\nonumber\\
&\times&
\frac{\Delta}{2i\hbar}\left(\frac{\beta_0(t)}{\beta_0^*(t)}+\frac{\beta^*_0(t)}{\beta_0(t)}\right)
\frac{1}{-2i\frac{E_F}{\hbar}+2i\omega_c\beta^*_0(t)\beta_0(t)}
\left(1-e^{\left(2i\frac{E_F}{\hbar}-2i\omega_c\beta^*_0(t)\beta_0(t)\right)t}\right).
\end{eqnarray}
\end{widetext}
This means that in the limit $\hbar\omega_c\ll\Delta\ll E_F$, over a time interval $\mathcal{O}(\hbar/\Delta)$, the wavepacket is appreciably
Andreev scattered only if
\begin{eqnarray}
\left|\frac{E_F}{\Delta}-\frac{\hbar\omega_c}{\Delta}\beta^*_0(t)\beta_0(t)\right|\lesssim 1
\end{eqnarray}
and if the d-wave form factor amplitude
\begin{eqnarray}
\left(\frac{\beta_0(t)}{\beta_0^*(t)}+\frac{\beta^*_0(t)}{\beta_0(t)}\right)
\end{eqnarray}
is maximized i.e. the wavepacket is located in the anti-node.
Otherwise, the wavepacket continues moving along the constant ($\Delta=0$) energy contours, essentially as if the system was a normal metal.

\bibliography{biblio}
\end{document}